\documentclass[journal]{IEEEtran}

\usepackage{cite}
\usepackage{url}

\usepackage{textcomp}
\usepackage{xcolor}
\usepackage{hyperref}

\usepackage{booktabs}	
\usepackage{diagbox}    
\usepackage{multirow}   

\usepackage{verbatim}	

\usepackage{amsmath,amssymb,amsfonts,graphicx}
\usepackage{epstopdf}

\newtheorem{definition}{Definition}[section]

\usepackage{graphicx,epstopdf,algpseudocode,caption,url}   
\usepackage[ruled,linesnumbered]{algorithm2e}
\usepackage{amssymb}
\usepackage{hyperref}

\ifCLASSINFOpdf

\else

\fi

\begin{document}

\title{Towards Target High-Utility Itemsets}

\author{Jinbao Miao, Wensheng Gan*, Shicheng Wan,  Yongdong Wu, Philippe Fournier-Viger	

	\thanks{This work was supported in part by the National Natural Science Foundation of China (Grant No. 62002136), Natural Science Foundation of Guangdong Province of China (Grant No. 2022A1515011861), Guangzhou Basic and Applied Basic Research Foundation (Grant No. 202102020277). (Corresponding author: Wensheng Gan)}

	\thanks{Jinbao Miao, Wensheng Gan, and Yongdong Wu are with the College of Cyber Security, Jinan University, Guangzhou 510632, China. (E-mail: osjbmiao@gmail.com, wsgan001@gmail.com, wuyd007@qq.com)}
	
	\thanks{Shicheng Wan is with the School of Computer Science and Technology, Guangdong University of Technology, Guangzhou 510006, China. (E-mail: scwan1998@gmail.com)}

	\thanks{Philippe Fournier-Viger is with the College of Computer Science and Software Engineering, Shenzhen University, China. (E-mail: philfv@qq.com)}

}


\maketitle

\begin{abstract}
	
For applied intelligence, utility-driven pattern discovery algorithms can identify insightful and useful patterns in databases. However, in these techniques for pattern discovery, the number of patterns can be huge, and the user is often only interested in a few of those patterns.  Hence, targeted mining of high-utility itemsets has emerged as a key research topic, where the aim is to find a subset of patterns that meet a target pattern constraint instead of all patterns. This is a challenging task because efficiently finding tailored patterns in a very large search space requires a targeted mining algorithm. A first algorithm called TargetUM has been proposed, which adopts an approach similar to post-processing using a tree structure, but the running time and memory consumption are unsatisfactory in many situations. In this paper, we address this issue by proposing a novel list-based algorithm with pattern matching mechanism, named THUIM (Targeted High-Utility Itemset Mining), which can quickly match high-utility itemsets during the mining process to select the target patterns.  Extensive experiments were conducted on different datasets to compare the performance of the proposed algorithm with state-of-the-art algorithms. Results show that THUIM performs very well in terms of runtime and memory consumption, and has good scalability compared to TargetUM.

\end{abstract}


\IEEEpeerreviewmaketitle

\section{Introduction}
\label{sec:introduction}

Frequent pattern mining (FPM) \cite{luna2019frequent,luna2018apriori,gan2019survey,wu2020netncsp} is a research topic that has been well-studied in the last decades. FPM techniques focus on counting the support (number of occurrences) of patterns in a database to identify frequent ones. The mining result is not only intuitive, but also easy to understand. Hence, it has been applied to different kinds of databases, like quantitative transaction databases \cite{agrawal1994fast,han2000mining}, streaming databases \cite{lin2005mining}, and time series databases \cite{han1999efficient}. However, only considering the support metric brings the problem that all items (data elements) in a pattern are considered as having the same relative importance. To address this issue, weighted pattern mining (WPM) was proposed \cite{cai1998mining, lan2014efficient, bui2021mining, yun2006wspan}. In quantitative transaction databases, the concept of weights refers to some important factors, such as the unit profits of items. In this framework, even if some items have a low support, they  are still viewed as valuable if they have large weights. Nevertheless,  WPM techniques do not take the quantities of items into account. Thereby, it still cannot satisfy some practical needs. For example, market retailers are strongly interested in discovering itemsets that yield a high revenue, which depends on both unit prices and purchase quantities.

In view of this, a generic and superior framework named high-utility itemset mining (HUIM) \cite{liu2005two,tseng2010up} was proposed. Each pattern has an internal utility (e.g., purchase quantity) and external utility (e.g., unit price). The utility metric attracts a lot of interest for two reasons: compared with FPM, UPM (Utility Pattern Mining) has the advantages of WPM; and compared with WPM, it considers the importance of distinct items more accurately. An itemset is called a high-utility itemset (abbreviated as HUI) if its real utility (importance) is no less than a user-specified minimum utility (simplified as \textit{minUtil}) threshold; otherwise, it is a low-utility itemset, which is uninteresting. After more than a decade of development, mining HUIs in quantitative transaction databases has been utilized in many applications such as user behavior analysis \cite{shie2013mining}, website click-stream analysis \cite{yun2018damped,shie2010online}, biomedical analysis \cite{gan2021survey}, and cross-marketing analysis \cite{tseng2012efficient, yen2007mining}. In general, discovering HUIs is a difficult task since the \textit{downward-closure} property \cite{agrawal1994fast} does not hold. In other words, it is not guaranteed that subsets of an HUI are HUIs. Another challenge is how to effectively prune the search space and efficiently capture all HUIs without missing any, for a very large search space, especially when databases contain lots of long transactions.  As a solution to the first problem, Liu \textit{et al.} \cite{liu2005two} proposed an overestimation upper-bound named transaction weighted utilization (abbreviated as \textit{TWU}). For brevity, this new upper-bound satisfies the \textit{downward-closure} property, which means a low-\textit{TWU} itemset cannot have high-\textit{TWU} (and also high-utility) supersets. For the second issue, many researchers have designed improved utility data mining algorithms. Many utility mining studies \cite{chen2021topic, lin2016fhn, liu2012mining, tseng2012efficient, zida2017efim} apply the \textit{TWU} to facilitate the mining process and use efficient data structures and pruning strategies to improve the performance of algorithms. There are already several algorithms in the high-utility itemset mining (HUIM) literature, such as list-based (i.e., HUI-Miner \cite{liu2012mining}, FHM \cite{fournier2014fhm}, FHN \cite{lin2016fhn}, and HUOPM \cite{gan2020huopm}),  tree-based (i.e., IHUP \cite{ahmed2009efficient}, UP-Growth \cite{tseng2012efficient} and MU-Growth \cite{yun2014high}), and projection-based (i.e., EFIM \cite{zida2017efim}) algorithms.

However, intuitively, most of HUIM algorithms aim to discover as much hidden information as possible from databases. However, in some cases, the requirements are the opposite. For instance, in the marketing analysis field, retailers are interested in goods which yield a high profit, and hence to find as much HUIs as possible. On the contrary, customers are more concerned about only some of the HUIs because of affordability and interest. Therefore, mining user-specified itemsets was proposed as a new interesting task, named targeted utility mining (abbreviated as TaUM) \cite{miao2021targetum,miao2022targetum}. To our best knowledge, there are some studies about target-based pattern mining such as target-oriented frequent itemset querying \cite{shabtay2018guided}, target-based association rule mining \cite{abeysinghe2017query, fournier2013meit} and sequential pattern querying \cite{chand2012target, chueh2010mining, zhang2021tusq}. However, there is little information in the literature about TaUM. This paper  focuses on the problem of querying special HUIs which contain target itemsets. For the TaUM task, the state-of-the-art algorithm is TargetUM \cite{miao2021targetum,miao2022targetum}, which utilizes the utility-list data structure and a querying trie to discover target high-utility itemsets. Nevertheless, the TargetUM algorithm faces the problem of high memory consumption. This is in part because items with low \textit{TWU} values are considered for generating high-level itemsets.

To address these issues, we propose a novel algorithm named THUIM (Targeted High-Utility Itemset Mining) as well as a compact data structure for efficiently mining the target HUIs in quantitative transaction databases. The major contributions of this study are:

\begin{itemize}
	\item  We first define the concept of pattern matching mechanism for targeted mining/searching interesting patterns. To the best of our knowledge, we are the first to propose a one-phase algorithm to address the very challenging problem of targeted utility mining.

	\item  The utility-list is extended as a new data structure, which is utilized for targeted mining of high-utility itemsets without scanning the database multiple times.

	\item We propose an efficient one-phase algorithm, namely THUIM, that can significantly reduce the search space using the matching mechanism.

	\item Extensive experiments were done on real and synthetic datasets. Results show that the proposed THUIM algorithm is efficient for the problem of targeted mining of high-utility itemsets in large-scale databases.
\end{itemize}

The rest of the paper is organized as follows. Related work is described in Section \ref{sec:relatedwork}. Some key preliminaries are introduced in Section \ref{sec:preliminaries}. The details of the proposed algorithm are presented in Section \ref{sec:algorithm}. Then, extensive experiments and a comprehensive analysis of results are presented in Section \ref{sec:experiments}. Finally, a conclusion is drawn and future work is discussed in Section \ref{sec:conclusion}.

\section{Related Work}  \label{sec:relatedwork}

In this section, we briefly review some studies about frequency-based itemset mining (FIM), utility-based itemset mining (UIM), and targeted itemset mining methods.

\subsection{Frequency-based itemset mining}

In the FIM domain, Apriori \cite{agrawal1994fast} is the most famous level-wise algorithm. It first generates low-level (short) candidates, and then iteratively constructs high-level (longer) frequent itemsets. Apriori applies the Apriori property (also known as downward-closure property) to reduce the search space, which states that an infrequent itemset cannot be a subset of a frequent itemset. Generating too many candidates and scanning the database many times are obvious drawbacks of Apriori that lead to poor performance in some situations. Some studies proposed improvements \cite{liu2005two,pasquier1999discovering,szathmary2007towards} to address these issues. Han \textit{et al.} \cite{han2000mining} proposed a tree-based FIM algorithm named FP-Growth, using a compact and extended prefix-tree structure named FP-tree. The FP-Growth algorithm reorganizes all the data from  a database in support descending order and insert it into an FP-tree, which makes the data highly compressed and eliminates the need of scanning the database multiple times. In recent years, some researchers have done additional improvements \cite{aryabarzan2018negfin,luna2019frequent} because FIM remains a popular problem. However, most of the FIM algorithms have the critical shortcoming that the frequency metric does not capture the relative importance of items, such that for example, diamonds brings a higher profit than a pen, although they have a lower selling frequency. To address this problem of FIM, Cai \textit{et al.} \cite{cai1998mining} proposed using weights to measure the relative importance of each item. In this framework, items have weights,  and even if some items are infrequent, they can also be discovered if they have large weights. After that, several weighted itemset mining studies \cite{lan2014efficient,yun2008efficient,bui2021mining} were done, and it has become an important variant of FIM.

\subsection{Utility-based itemset mining}

Although the weighted pattern mining (WPM) framework \cite{yun2006wspan,bui2021mining} has some applications, weights remain a solution that is often too simple and cannot fully capture the importance of patterns. For instance, WPM cannot address the requirements of users who are interested in mining itemsets that yield a high profit. In view of this, high-utility itemset mining (HUIM) \cite{gan2021survey} has emerged as an important topic in the last decades. Utility represents how much an item is profitable in quantitative transaction databases, and it has good interpretability. If the utility of an itemset is no less than a user-specified minimum utility \textit{minUtil} threshold, it is deemed to be a high-utility itemset (abbreviated as HUI); otherwise, it is said to be a low-utility itemset, which is uninteresting. Discovering HUIs in a database is not an easy task since the \textit{downward-closure} property \cite{agrawal1994fast} in FIM does not hold in utility mining. It is difficult to estimate if the supersets of a HUI are HUIs or not. A naive approach is to count the utilities of all itemsets by scanning the database, and then keep the HUIs. While obviously, this approach suffers from the problem of a very large search space, especially for long transactions. Hence, Liu \textit{et al.} \cite{liu2005two}  proposed the transaction-weighted utilization (\textit{TWU}) to solve this issue and an algorithm. This latter first calculates the \textit{TWU} of all items by scanning the database, and then keep candidates that may be HUIs. In the second step, the algorithm identifies the real HUIs from those candidates with an additional database scan. Generally, we can classify early HUIM algorithms such as \cite{ahmed2009efficient, han2000mining,liu2005two, tseng2012efficient} as two-phase-based algorithms.

Although two-phase algorithms effectively reduce the search space and efficiently find the complete set of HUIs for HUIM, runtime and memory consumption remain a major issue. As an alternative to two-phase-based algorithms, single-phase-based algorithms have emerged as an important topic. HUI-Miner \cite{liu2012mining} is the first work for mining HUIs without candidate generation. It utilizes a compact utility-list structure to avoid scanning a database repeatedly.  An intersection operation for low-level utility-lists allows obtaining the utility information of all HUIs. Many list-based algorithms have been developed for HUIM such as FHM \cite{fournier2014fhm} and FHN \cite{lin2016fhn}. The worst shortcoming of list-based algorithms is that lists may consume much memory when dealing with huge databases containing thousands of items. Zida \textit{et al.} \cite{zida2017efim} proposed the EFIM algorithm, which is inspired by the LCM algorithm \cite{uno2004lcm} for FIM. EFIM adopts a depth-first search mechanism to process each itemset in linear time and space. Moreover, it integrates a novel array-based utility counting technique to calculate two new upper bounds (\textit{local-utility} and \textit{subtree-utility}) to reduce search space effectively. With these efficient techniques, EFIM is in general two to three orders of magnitudes faster than previous work.  TOPIC \cite{chen2021topic} can handle itemsets with negative utility and find top-$k$ patterns to cope with the problem of setting a suitable \textit{minUtil} value. Up to now, there are many other advanced studies of HUIM, such as  \cite{mai2017lattice,wu2019high,song2021generalized}.

\subsection{Targeted-based itemset mining}

In the fields of FIM and HUIM, discovering all the interesting itemsets in quantitative transaction databases is a vital task. However, in real applications, customers often prepare a bargain list before shopping that contains only a few items, which can effectively help them save time. In another situation, shareholders usually need to know when it is a suitable time to sell the stocks that they own at an acceptable price but are less interested in other stocks. Based on such observations, Kubat \textit{et al.} \cite{kubat2003itemset}  proposed a target-oriented querying task for FIM \cite{kubat2003itemset} and designed algorithms to answer targeted queries with an Itemset-Tree (IT) structure. That structure can be incrementally updated and efficiently queried to find itemsets that meet some constraints. However, the memory consumption of the IT can be huge for large incremental  databases. Thus, a new efficient data structure named Memory Efficient Itemset Tree (MEIT) \cite{fournier2013meit} was proposed. The MEIT structure compresses the tree nodes and incorporates a decompression mechanism to reduce memory consumption. Extensive experiments show that the IT is up to 60\% bigger than the MEIT.  Miao \textit{et al.} \cite{miao2021targetum} proposed a work, namely TargetUM, on incorporating the concept of utility into targeted HUIM. As a trie-based algorithm, TargetUM is a generic framework as stated above. As the first solution to the targeted HUIM problem, TargetUM brought forward some novel ideas. Nevertheless, on the one hand, TargetUM inherits the limitations of list-based algorithms; and on the other hand, it  can take many unpromising items into account to find the final result, which can impair performance. To handle sequence data, targeted sequential pattern mining was first introduced recently \cite{huang2022taspm}. Zhang \textit{et al.} \cite{zhang2021tusq} designed a framework for quickly querying high-utility sequences. That is the first study to incorporates the utility into target-oriented sequence mining. Currently, there remain much work to design more efficient targeted utility mining algorithms. In fact, TargetUM \cite{miao2021targetum,miao2022targetum} is the only algorithm for utility-driven targeted itemset querying. This is an unsatisfactory result, which also motivates us to design a more efficient approach.

\section{Preliminaries and Problem Statement}
\label{sec:preliminaries}

This section first introduces some preliminary definitions about HUIs. The notations and definitions are mostly based on those of prior studies \cite{miao2021targetum,miao2022targetum, zida2017efim}. The new algorithm THUIM will be introduced in the next section.

\subsection{Preliminaries}

Let there be a set $I$ = \{$x_1$, $x_2$, $x_3$, $\ldots$, $x_n$\} of $n$ distinct items in a database $\mathcal{D}$. And $\mathcal{D}$ is a set of $m$ transactions, which is defined as \{$T_1$, $T_2$, $T_3$, $\ldots$, $T_m$\}. We assume that each transaction $T_{tid}$ has a unique ID \textit{tid}, and a transaction consists of a finite set of items (\{$x_1$, $x_2$, $\ldots$, $x_p$\}). An itemset $X$ is a subset of $I$ and is called an $l$-itemset if it has $l$ (1 $\le$ $\mid l \mid$ $\le$ $n$) distinct items. The utility of an item $x_i$ (1 $\le$ $i$ $\le$ $n$) consists of two aspects: 1) the internal utility (i.e., quantity),  denoted as $q(x_i, T_j)$, which is the quantity associated with $x_i$ in the transaction $T_j$ (1 $\le$ $j$ $\le$ $m$); 2) the external utility (i.e., unit profit), defined as $p(x_i)$, which is a positive value, for $x_i$ in $\mathcal{D}$. A simple quantitative transaction database is given in Table \ref{tab:db_example}, which is composed of seven transaction data, and include seven items \{$a$, $b$, $c$, $d$, $e$, $f$, $g$\}. The external utilities of these items are $p(a)$ = \$3, $p(b)$ = \$5, $p(c)$ = \$1, $p(d)$ = \$5, $p(e)$ = \$3, $p(f)$ = \$4, and $p(g)$ = \$2, respectively.

\begin{table}[!h]	
	\centering
	\caption{A quantitative transaction database.}
	\label{tab:db_example}
	\begin{tabular}{ccc}
		\hline
		\textbf{TID} & \textbf{Transaction} & \textbf{Quantity}  \\ \hline
		$T_1$    	 & \{$d$, $e$, $f$\}	& \{2, 5, 6\}	\\
		$T_2$    	 & \{$a$, $b$, $d$, $f$\}	 & \{3, 8, 7, 1\}	\\
		$T_3$    	 & \{$a$, $b$, $e$, $f$\}    	& \{5, 3, 4, 3\}	\\
		$T_4$    	 & \{$a$, $b$, $c$, $d$, $f$\} 	   & \{4, 6, 1, 4, 2\}  \\
		$T_5$    	 & \{$b$, $d$, $e$, $f$, $g$\}     & \{5, 3, 7, 6, 3\}	\\
		$T_6$    	 & \{$a$, $b$, $d$, $f$, $g$\}     & \{8, 7, 2, 1, 1\}	\\
		$T_7$    	 & \{$a$, $b$, $c$, $e$, $f$, $g$\} 	& \{1, 1, 6, 5, 4, 2\}	\\
		\hline
	\end{tabular}
\end{table}

\begin{definition}[utility of itemset]
	\rm Given an item $x$, the utility of $x$ in a transaction $T_{tid}$ is denoted as $U(x, T_{tid})$ = $q(x, T_{tid})$ $\times$ $p(x)$ (where $q(x, T_{tid})$ represents the number of occurrences in $T_{tid}$, that is the internal utility). The real utility of $x$ in a database $\mathcal{D}$ is the sum of the utility values, denoted as $U(x)$ = $\sum_{x \in T_{tid} \land T_{tid} \in \mathcal{D}}$$(q(x, T_{tid})$ $\times$ $p(x))$. Then, the utility of an itemset $X$ is defined as $U(X)$ = $\sum_{x_i \in X \wedge X \subseteq T_{tid}}$$U(x_i, T_{tid})$.
\end{definition}

For example, in Table \ref{tab:db_example}, $U(e, T_1)$ = $q(e, T_1)$ $\times$ $p(e)$ = 5 $\times$ \$3 = \$15, $U(e, T_3)$ = \$12, $U(e, T_5)$ = \$21 and $U(e, T_7)$ = \$15. Therefore, the utility of item $e$ in $\mathcal{D}$ is $U(a)$ = $T_1$ + $T_3$ + $T_5$ + $T_7$ = \$15 + \$12 + \$21 + \$15 = \$63. Setting itemset $X$ = $\{f, g\}$, $U(X)$ = $U(f)$ + $U(g)$ = $q(f, T_5)$ $\times$ $p(f)$ + $q(f, T_6)$ $\times$ $p(f)$ + $q(f, T_7)$ $\times$ $p(f)$ + $q(g, T_5)$ $\times$ $p(g)$ + $q(g, T_6)$ $\times$ $p(g)$ + $q(g, T_7)$ $\times$ $p(g)$ = \$24 + \$4 + \$16 + \$6 + \$2 + \$4 = \$56. And if $X$ = $\{e, g\}$, then $U(\{e, g\})$ = \$27.

\begin{definition}[transaction-weighted utilization \cite{liu2012mining}]
	\rm The utility of a transaction ($T_{tid}$ $\in$ $\mathcal{D}$) is the sum of the utility of all items contained in it, denoted as $TU(T_{tid})$ = $\sum_{x_i \in T_{tid}}U(x_i$, $T_{tid})$. The transaction-weighted utilization of itemset $X$ is defined as \textit{TWU}($X$) = $\sum_{X \subseteq T_{tid} \land T_{tid} \in \mathcal{D}}TU(T_{tid})$. 
\end{definition}

\textit{TWU} is a useful upper bound and can help cut off unpromising items in advance. Due to the space limitation, the proof and details can be found in Ref. \cite{liu2012mining}. For example, in Table \ref{tab:db_example}, $TU(T_1)$ = $U(d, T_1)$ + $U(e, T_1)$ + $U(f, T_1)$  = 2 $\times$ \$5 + 5 $\times$ \$3 + 6 $\times$ \$4 = \$49. Therefore, \textit{TWU}$(a)$ = $TU(T_2)$ + $TU(T_3)$ + $TU(T_4)$ + $TU(T_6)$ + $TU(T_7)$ = \$88 + \$54 + \$71 + \$75 + \$49 = \$337, and also we have \textit{TWU}$(e)$ = \$49 + \$54 + \$91 + \$49 = \$243.

\begin{definition}[high-utility itemset \cite{liu2005two,gan2021survey}]
	\rm The minimum utility threshold (abbreviated as \textit{minUtil}) is a user-specified parameter. If the utility of an itemset $X$ is no less than \textit{minUtil}, where $U(X)$ $\ge$ \textit{minUtil}, we call $X$ a high-utility itemset (abbreviated as HUI); otherwise, it is a low-utility itemset, which we are not interested in. In this paper, \textit{minUtil} ($\sigma$) is used to discover target HUI directly.
\end{definition}

\begin{definition}[target high-utility itemset \cite{miao2021targetum,miao2022targetum}]
	\rm The target itemset is a user-specified set of items $T^\prime$ = $<$$x_1$, $x_2$, $\dots$, $x_i$, $\dots$, $x_n$$>$ ($i \in$ $[1, n]$ $\land x_i$ $\in I)$ which the user(s) are interested in. It should be pointed out that $\mid$$T^\prime$ $\mid$ $\ge$ 1. Given a high utility itemset $X$, and some target itemset $T^\prime$, if $T^\prime$ $\subseteq$ $X$, we say that  $X$ is a target high-utility itemset (abbreviated as \textit{THUI}). 
\end{definition}

\subsection{Problem statement}

In general, HUIM algorithms aim to find a complete set of HUIs in quantitative transaction databases. However, in real applications, customers always need a part of those HUIs, and sometimes they may have different requirements at different times. For example, a user query may seek all HUIs containing milk and bread together; another query may seek all HUIs containing milk or bread alone. 

Based on Table \ref{tab:db_example}, if we set $T^\prime$ = $\{e, f\}$ and $\sigma$ = \$130, then we can get the THUIs: \{$e$, $b$, $f$\} (= \$145) and \{$e$, $f$\} (= \$139). When setting $T^\prime$ = \{$c$, $f$\} and $\sigma$ = \$50, we can obtain \{$c$, $a$, $d$, $b$, $f$\} (= \$71), \{$c$, $a$, $b$, $f$\} (= \$81), \{$c$, $d$, $b$, $f$\} (= \$59), and \{$c$, $b$, $f$\} (= \$66). Based on the above  description, we give the problem statement.

\textbf{Problem statement}: The targeted high-utility itemset mining task aims to discover all THUIs for a given \textit{minUtil} threshold value, without scanning the database multiple times. In this paper, the main task of THUI mining is to directly get all the itemsets that contain the target item(s) and have a utility that is no less than $\sigma$. 

\section{The THUIM Algorithm}
\label{sec:algorithm}

It was observed that the prior TargetUM algorithm has relatively poor performance, and can have long runtime and high memory consumption for dense as well as large datasets. Therefore, a more efficient algorithm is essential for the targeted pattern mining problem. In this section, we propose the THUIM algorithm, which is based on the HUI-Miner algorithm \cite{liu2012mining}. By integrating an efficient targeted pattern matching mechanism in HUI-Miner, targeted high-utility itemsets can be mined directly. THUIM is a more efficient than TargetUM as it will be demonstrated.

\subsection{Matching mechanism}

The HUI-Miner algorithm uses the utility and  remaining utility as judgment conditions to reduce the search space, which can quickly and efficiently mine all high-utility itemsets. For this reason, we introduce a matching mechanism based on HUI-Miner in the proposed  THUIM algorithm. Assume a total order $\prec$ on items, that is the \textit{TWU} ascending order. That is to say for any two items $x_i$ and $x_j$ if $x_j$ is after $x_i$ we have \textit{TWU}($x_j$) $\ge$ \textit{TWU}($x_i$). For example, from Table \ref{Tab:twu}, we have \textit{TWU}($a$) $\ge$ \textit{TWU}($c$), and after sorting, $a$ is after $c$.

\begin{table}[!h]
	\begin{center}
		\caption{Transaction-weighted utility.}
		\label{Tab:twu}
		\begin{tabular}{lccccccc}
			\hline
			\textbf{item} 		  & $a$  & $b$  & $c$   & $d$  & $e$  & $f$  & $g$  \\ \hline
			\textbf{\textit{TWU}} & \$337 & \$428 & \$120 & \$374 & \$243 & \$477 & \$215 \\ \hline
		\end{tabular}
	\end{center}
\end{table}

\begin{definition}[serial number on items]
	\label{def:sn}
	\rm According to the \textit{TWU}-ascending order, a unique serial number is assigned to each item. More precisely, consecutive integer serial numbers are assigned to all items having a \textit{TWU} that is no less than $\sigma$ in increasing order starting from 1, denoted as $sn$($x_i$) that is $x_i$ $\in$ $I$ $\land$ \textit{TWU}($x_i$) $\ge$ $\sigma$.
\end{definition}

For the running example, we  get that \textit{TWU}($c$) = \$120 $<$ \textit{TWU}($g$) = \$215 $<$ \textit{TWU}($e$) = \$243 $<$ \textit{TWU}($a$) = \$337 $<$ \textit{TWU}($d$) = \$374 $<$ \textit{TWU}($b$) = \$428 $<$ \textit{TWU}($f$) = \$477. Thus, the total order $\prec$ on items is $c$ $\prec$ $g$ $\prec$ $e$ $\prec$ $a$ $\prec$ $d$ $\prec$ $b$ $\prec$ $f$. If we set $\sigma$ = \$130, we have $sn$($g$) = 1, $sn$($e$) = 2, $sn$($a$) = 3, $sn$($d$) = 4, $sn$($b$) = 5, and $sn$($f$) = 6. The matching mechanism in THUIM is implemented using the serial numbers. According to \textbf{Definition} \ref{def:sn}, for each item $x_i$ such that \textit{TWU}($x_i$) $\ge$ $\sigma$, there is a unique serial number. This is to be able to distinguish between items in the case where multiple items have the same \textit{TWU}. To facilitate pattern matching, preprocessing is required for the target pattern $T^\prime$, so that it is in line with the overall order of the algorithm that is the \textit{TWU}-ascending order.

\begin{definition}[matching]
	\rm For two itemsets $X$ and $Y$, if $Y$ is a subsequence of $X$, that is to say $Y$ $\subseteq$ $X$, then $Y$ matches with $X$ and $X$ is matched by $Y$. If $Y$ $\subsetneq$ $X$, then $Y$ and $X$ mismatch, and if $Y$ is partly contained in $X$ and the other part is indeterminate, then $Y$ and $X$ can be matched.
\end{definition}

For example, consider the itemsets $X$ = \{$e$, $b$, $f$\} and $Y$ = \{$b$, $f$\}, $Y$ $\subseteq$ $X$, thus $X$ is matched by $Y$. Thus, if the user sets $T^\prime$ = $\{e, f\}$,  $X$ is  matched by $T^\prime$.

\subsection{Utility-list structure}

\begin{figure*}[!htbp]
	\centering
	\includegraphics[scale=0.5]{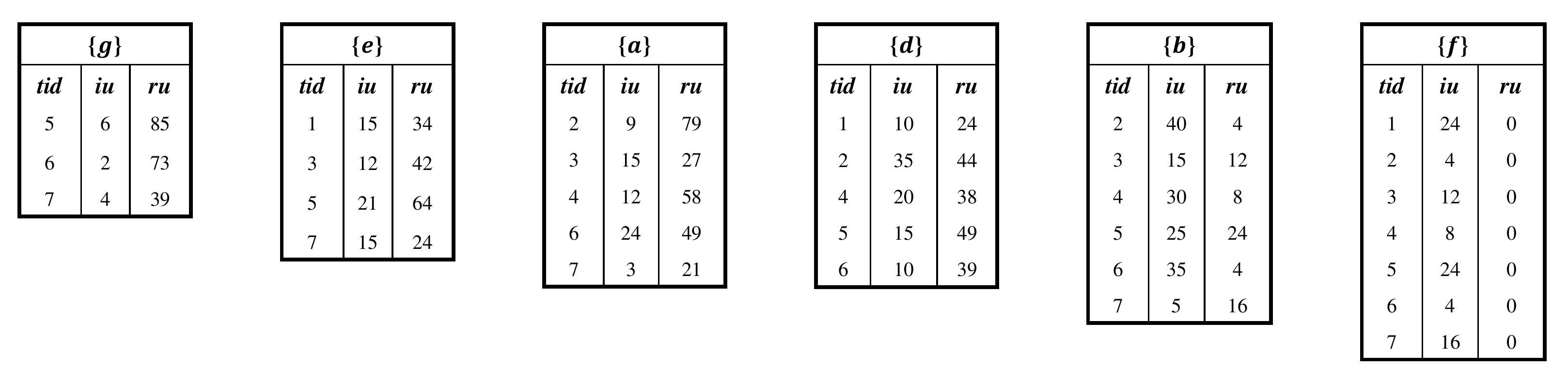}
	\caption{The \textit{UList}s of 1-itemsets for $T^\prime$ = $\{e, f\}$.}
	\label{fig:1UL}
\end{figure*}

The HUI-Miner algorithm mainly uses the utility-list data structure to mine the full set of HUIs \cite{liu2012mining}. In the THUIM algorithm, the utility-list is still a very important component. The utility-list contains multiple tuples, each tuple containing three fields which are the id of the transaction (\textit{tid}) containing the itemset $X$, the utility of $X$ in that transaction, and the remaining utility \cite{liu2012mining} of $X$ in the current transaction, that is (\textit{tid}, $iu$, $ru$). A utility-list \textit{UList} has the form $<$($tid_1$, $iu_1$, $ru_1$), ($tid_2$, $iu_2$, $ru_2$), $\dots$, ($tid_q$, $iu_q$, $ru_q$)$>$. All the utility-lists of 1-itemsets  are shown in Fig. \ref{fig:1UL}. 

\begin{figure*}[!htbp]
	\centering
	\includegraphics[scale=0.5]{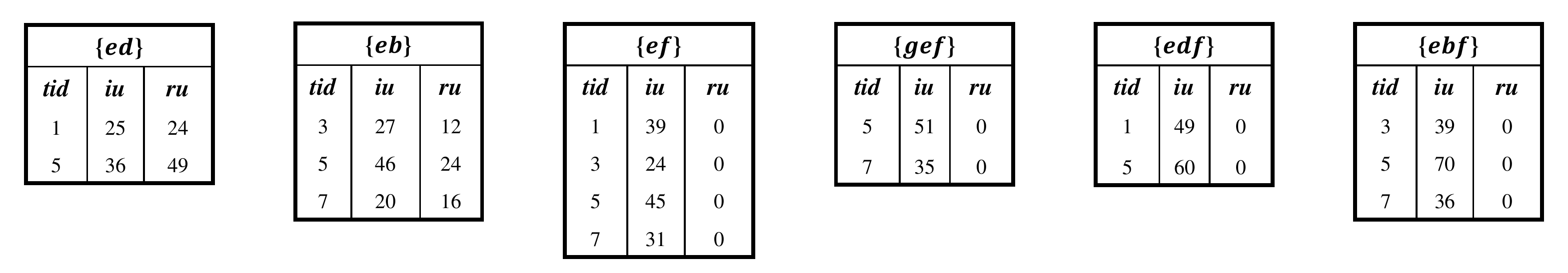}
	\caption{Some \textit{UList}s of $k$-itemsets with $T^\prime$ = $\{e, f\}$.}
	\label{fig:kUL}
\end{figure*}

The ($k$+1)-itemset(s) can be obtained from the $k$-itemset(s) by merging the \textit{ULists}. Suppose the utility-list of itemset $X_a$ is \textit{ULa} and the utility-list of the itemset $X_b$ is \textit{ULb}, when joining two \textit{UList} (\textit{ULa} and \textit{ULb}), the first step is to ensure that the same transaction \textit{tid} exists in \textit{ULa} and \textit{ULb}. Then, sum the utility and take the remaining utility of the item corresponding to the largest \textit{TWU}. Note that when summing the utility, the utility corresponding to the common prefix should be subtracted, but for a 1-itemset, there is no common prefix. That is for two $k$-itemsets $X_a$ and $X_b$ having a common prefix $X_c$, we can get the ($k$+1)-itemset $X_{ab}$, having $iu$($X_{ab}$, \textit{tid}) = $iu$($X_a$, \textit{tid}) + $iu$($X_b$, \textit{tid}) ($k$ = 1) and $iu$($X_{ab}$, \textit{tid}) = $iu$($X_a$, \textit{tid}) + $iu$($X_b$, \textit{tid}) - $iu$($X_c$, \textit{tid}) ($k$ $\ge$ 2). The specific details can be seen in \textbf{Algorithm} \ref{algo:construct}. And the \textit{UList} of $k$-itemsets ($k$ $\ge$ 2) are shown in Fig. \ref{fig:kUL}.

\subsection{Efficient pruning strategies}

In TargetUM \cite{miao2021targetum,miao2022targetum}, when performing an upward query, the target pattern $T^\prime$ and the high-utility itemsets are matched one by one. And the item's \textit{TWU} determines whether the target item can be successfully matched. However, different datasets have variability, i.e., different items may have the same \textit{TWU}. Thus, it is also necessary to specifically determine whether items are the same when the \textit{TWU} is the same. To facilitate comparison, we introduce the concept of \textit{serial number} (\textbf{Definition} \ref{def:sn}). The pruning strategy is designed by \textit{serial number} to greatly improve the efficiency of THUIM for mining the target high-utility itemsets.

\textbf{Pruning strategy}: when extending the search for a target high-utility itemset $X$, if the serial number of the extended item is larger than that of the current item to be matched in $T^\prime$, it means that there is a mismatch, which means that $T$ must not be included in $X$. That is said for two items $x_t$ ($x_t$ $\in$ $T^\prime$) and $x_i$ ($x_i$ $\in$ $X$ $\land$ $i$ == $|$$X$$|$) which need to be matched, if $sn$($x_i)$ $> sn$($x_t$), then the search will be terminated, and the itemsets that follow must not contain $T^\prime$.

\textbf{Proof}: All the single items and $T^\prime$ are sorted in \textit{TWU} ascending order, thus according to \textbf{Definition} \ref{def:sn}, if \textit{TWU}($x_i$) $>$ \textit{TWU}($x_j$), we have $sn$($x_i$) $> sn(x_j)$. For the item $x_t$ in $T^\prime$ (0 $\le$ $t$ $<$ $|$$T^\prime$$|$), if $sn(x_i)$ == $sn(x_t)$, it indicates a successful match, and the next recursion will match $x_{i+1}$ and $x_{t+1}$. If $sn(x_i)$ $\neq$ $sn(x_t)$, the match fails and the next recursion will match $x_{i+1}$ and $x_t$. Thus, if $sn(x_i)$ $> sn(x_t)$, for any item $x_j$ ($j$ $>$ $i$), we will have $sn(x_j)$ $> sn(x_t)$.

\begin{figure}[!htbp]
	\centering
	\includegraphics[scale=0.4]{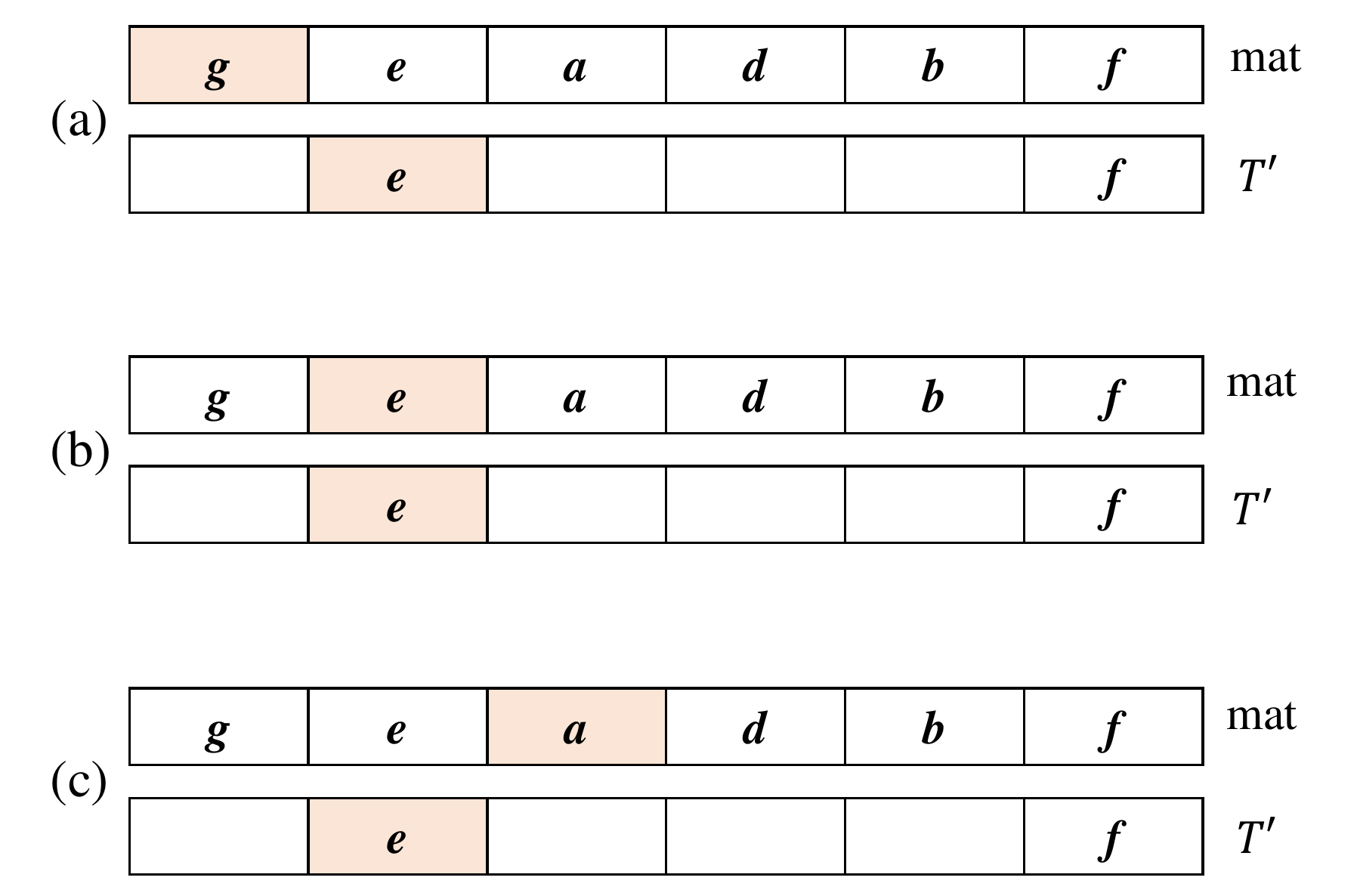}
	\caption{Match with serial number for $T^\prime$ = $\{e, f\}$.}
	\label{fig:strategy}
\end{figure}

For example, from Fig. \ref{fig:strategy}, we can get that $T^\prime$ = \{$e$, $f$\} and \textit{mat} which contains all the items that can be expanded. For the Fig. \ref{fig:strategy} (a), the items to be matched are $g$ (\textit{mat}) and $e$ ($T^\prime$), have $sn(g)$ $<$ $sn(e)$, next will match $e$ (\textit{mat}) and $e$ ($T^\prime$). Note that for the next item to be matched is uncertain, thus for $e$ ($T^\prime$), \textit{mat} may or may not be included. Considers Fig. \ref{fig:strategy} (b), have $sn(e)$ == $sn(e)$, and next will match $a$ (\textit{mat}) and $f$ ($T^\prime$). Finally, if the items can be matched as $a$ (\textit{mat}) and $e$ ($T^\prime$) in Fig. \ref{fig:strategy} (c), have $sn(a)$ $>$ $sn(e)$, and the extended match will be terminated. A concrete example will be shown in the algorithm section of this paper.

The utility-list is composed of multiple triples ($<$\textit{tid}, \textit{iu}, \textit{ru}$>$). Counting the sum of $iu$ of all tuples in a utility-list can get the exact utility of itemset $X$ (\textit{sumIutils}). Similarly, counting the sum of $ru$ of all tuples in the utility-list can get the maximum utility of itemset $X$ that can be extended (\textit{sumRutils}). Therefore, computing the sum of \textit{sumIutils} and \textit{sumRutils} yields an upper bound on utility ($\hat{\Theta}$). If $\hat{\Theta}$ $\ge$ $\sigma$, it is possible to obtain a high-utility itemset by extending the itemset $X$. Otherwise, it means that any high-utility itemsets cannot be obtained by extending the itemset $X$ \cite{liu2012mining}.

\subsection{Proposed targeted pattern querying algorithm}

In this subsection, the details of the proposed querying algorithm for mining target high-utility itemsets are discussed. Three main sub-functions are included, the data processing function (\textbf{Algorithm} \ref{algo:THUIM}), the itemset mining function (\textbf{Algorithm} \ref{algo:THUIMminer}), and the utility-list construction function (\textbf{Algorithm} \ref{algo:construct}).

\begin{algorithm}[h]
	\caption{THUIM ($\mathcal{D}$, $\sigma$, $T^\prime$) procedure}
	\label{algo:THUIM}
	\LinesNumbered
	scan the database $\mathcal{D}$ to get the basic data format and count the \textit{TWU} values of all 1-itemsets, i.e. \textit{TWU}($x_i$) ($x_i$ $\in$ $I$);
	
	scan $I$ to get the set  $I^\prime$ where $x_i$ $\in$ $I$ $\land$ \textit{TWU}($x_i$) $\ge$ $\sigma$;
	
	scan and evaluate $T^\prime$;
	
	\If{\textit{TWU}($x_i$) $<$ $\sigma$ (for all $x_i$ $\in$ $T^\prime$)}{
		return null;
	}
	
	sort  $I^\prime$ by the \textit{TWU} ascending order $\prec$;
	
	sort  $T^\prime$ by the \textit{TWU} ascending order $\prec$;
	
	scan  $I^\prime$ to get the order set of \textit{mapItemToOrder};
	
	second scan of  $\mathcal{D}$, and sort each transaction by the \textit{TWU} ascending order $\prec$ and construct the utility-list denoted as \textit{UList} for each item in $I^\prime$;
	
	\textbf{call} \textbf{\textit{THUIM}($\phi$, \textit{UList}, $\sigma$, 0, $|$$T^\prime$$|$, \textit{mapItemToOrder})}.
	
\end{algorithm}

Consider \textbf{Algorithm} \ref{algo:THUIM}, which has three input parameters, which are the database ($\mathcal{D}$), minimum utility threshold ($\sigma$) and target pattern ($T^\prime$). The database ($\mathcal{D}$) needs to be scanned twice, the first time to obtain the upper bound of the utility of all 1-itemsets (\textit{TWU}), and filter out the 1-itemsets that do not satisfy the minimum utility threshold ($\sigma$) to obtain the set $I^\prime$ (lines 1-2). The second database scan is to construct the utility-list of all 1-itemsets, and start the mining of target HUIs (lines 10-11). In \textbf{Algorithm} \ref{algo:THUIM}, the same preprocessing is needed to sort all the items in $T^\prime$ in ascending order according to the \textit{TWU}, and to map the set $I^\prime$ to obtain the serial number set \textit{mapItemToOrder} (lines 7-9). The \textit{mapItemToOrder} variable is a mapping structure, that is, renumbering all items in $I^\prime$. For example, take Table \ref{tab:db_example} and assume that $\sigma$ = \$130. We have $I^\prime$ = \{$a$, $b$, $d$, $e$, $f$, $g$\}, and then we get \textit{mapItemToOrder} as shown in the Table \ref{Tab:seri} after sorting.

\begin{table}[!h]	
	\begin{center}
		\caption{A sample \textit{mapItemToOrder}}
		\label{Tab:seri}
		\begin{tabular}{lcccccc}
			\hline
			\textbf{item} 		  & $g$  & $e$  & $a$   & $d$  & $b$  & $f$ \\ \hline
			\textbf{\textit{Serial Number}} & 1 & 2 & 3 & 4 & 5 & 6 \\ \hline
		\end{tabular}
	\end{center}
\end{table}
 
\begin{algorithm}[h]
	\caption{THUIM procedure}
	\label{algo:THUIMminer}
	\LinesNumbered
	
	\KwIn{\textit{UListOfP}: the utility-list of itemset $P$; \textit{UList}: the set of utility-lists of extensions of itemset $P$ ; the minimum utility threshold ($\sigma$); \textit{index}: points to the current item in $T^\prime$ that needs to be matched; $\beta$: the size of $T^\prime$; \textit{mapItemToOrder}: mapping of items to serial numbers.} 
	
	\KwOut{all the target high-utility itemsets (\textit{THUIs}).}

	\For {each element of utility-list $x$ in \textit{UList}}{
		
		initialize \textit{currentIndex} $\leftarrow$ \textit{index};
		
		initialize \textit{oriOrder} $\leftarrow$ 0;
		
		initialize \textit{patOrder} $\leftarrow$ 0;
		
		\If{\textit{currentIndex} $<$ $\beta$}{
			\textit{oriOrder} $\leftarrow$ the serial number of \textit{x.item} in \textit{mapItemToOrder};
			
			\textit{patOrder} $\leftarrow$ the serial number of $T^\prime$[\textit{index}] in \textit{mapItemToOrder};
			
			\If{\textit{oriOrder} == \textit{patOrder}}{
				
				\textit{currentIndex} increasing;
			}
		}
		
		\If{\textit{currentIndex} $<$ $\beta$ AND \textit{patOrder} $<$ \textit{oriOrder}}{
			
			\textbf{break};
			
		}
		
		\If{\textit{x.sumIutils} $\ge$ $\sigma$ AND \textit{currentIndex} $\ge$ $\beta$}{
			new \textit{THUIs} $\leftarrow$ \textit{THUIs} $\cup$ $x$;
		}
		
		\If{\textit{x.sumIutils} + \textit{x.sumRutils} $\ge$ $\sigma$}{
			
			new utility-list \textit{newUList} $\leftarrow$ \textit{null};
			
			\For{each element of utility-list $y$ after $x$ in \textit{UList}}{
				
				\textit{newUList} $\leftarrow$ \textit{newUList} $\cup$ \textbf{construct}(\textit{UListOfP}, $x$, $y$);
			}
		}
		
		\textbf{call} \textbf{\textit{THUIM}(\textit{UList}, \textit{newUList}, $\sigma$, \textit{currentIndex}, $|$$T^\prime$$|$, \textit{mapItemToOrder})}.
	}
\end{algorithm}

\begin{algorithm}[h]
	\caption{construct procedure}
	\label{algo:construct}
	\LinesNumbered
	\KwIn{\textit{UListOfP}, \textit{UListOfPX}, \textit{UListOfPY}: the utility-lists of itemset $P$, $PX$ and $PY$.}
	
	\KwOut{the utility-list of itemset \textit{PXY} (\textit{UListOfPXY}).}
	
	\textit{UListOfPXY} $\leftarrow$ \textit{null};
	
	\For{each element $ex$ in \textit{UListOfPX}}{
		
		$ey$ $\leftarrow$ use binary search to find an element $e$ such that \textit{e.tid} == \textit{ex.tid} in \textit{UListOfPY};
		
		\If{\textit{ex.tid} == \textit{ey.tid}}{
			
			\eIf{\textit{UListOfP} == \textit{null}}{
				
				new \textit{exy} $\leftarrow$ $<$\textit{ey.tid}, \textit{ex.iutils} + \textit{ey.iutils}, \textit{ey.rutils}$>$;
			}{
				
				$e$ $\leftarrow$ use binary search to find $ex$ from \textit{UListOfP};
				
				new \textit{exy} $\leftarrow$ $<$\textit{ey.tid}, \textit{ex.iutils} + \textit{ey.iutils} - \textit{e.iutils}, \textit{ey.rutils}$>$;
			}
			
			\textit{UListOfPXY} $\leftarrow$ \textit{UListOfPXY} $\cup$ \textit{exy};
		}
	}
	
	\textbf{return} \textit{UListOfPXY}
\end{algorithm}

\textbf{Algorithm} \ref{algo:THUIMminer} shows the main process of targeted utility mining, taking six parameters as input, which are two utility-lists \textit{UListOfP} and \textit{UList}, where \textit{UList} is an extension of \textit{UListOfP}, the minimum utility threshold ($\sigma$), the pointer of $T^\prime$ (\textit{index}), the size of $T^\prime$ and the serial number set \textit{mapItemToOrder}. \textbf{Algorithm} \ref{algo:THUIMminer} performs item matching for each recursive execution, which are \textit{x.item} and $T^\prime$[\textit{index}] (lines 1-4). Firstly, we need to get the serial number with matching items \textit{oriOrder} and \textit{patOrder}. And if \textit{oriOrder} equals \textit{patOrder}, it means that items \textit{x.item} and $T^\prime$[\textit{index}] can be matched, and then the pointer \textit{index} is moved back one position (lines 5-11). For the items to be matched, if \textit{patOrder} $<$ \textit{oriOrder}, the traversed itemset  is a mismatch with $T^\prime$ according to the pruning strategy, and the current recursion terminates (lines 12-14). If \textit{x.sumIutils} $\ge$ $\sigma$ and \textit{currentIndex} $\ge$ $\beta$, a new \textbf{THUI} can be discovered. And if \textit{x.sumIutils} + \textit{x.sumRutils} $\ge$ $\sigma$,  extended utility-lists are built. Finally, there is a recursive call to \textbf{Algorithm} \ref{algo:THUIMminer} (lines 15-24). \textbf{Algorithm} \ref{algo:construct} shows how to construct a utility-list. To construct a utility list from the utility lists of some itemsets $X$ and $Y$ to generate the utility list of $XY$, it must be ensured that $X$ and $Y$ have an element $e$ with the same \textit{tid}, that is to say, the intersection of $X$ and $Y$ is not the empty set. If there is no such element $e$, it means that the itemset $XY$ does not exist in the dataset. For implementation details, one can refer to HUI-Miner \cite{liu2012mining}.

In \textbf{Algorithm} \ref{algo:THUIMminer}, it is crucial to match \textit{oriPrder} and \textit{patOrder}. Assume that $\sigma$ = \$130 and $T^\prime$ = $\{e, f\}$. In Fig. \ref{fig:1UL}, if $g$ and $d$ are joined to generate $gd$, then we need to compare $d$ with the first item $e$ in $T^\prime$ (note that $T^\prime$ is sorted). By looking up \textit{mapItemToOrder} we get $sn(d)$ = 4 and $sn(e)$ = 2, thus have $sn(d)$ $>$ $sn(e)$. This is a mismatch because $e$ doesn't appear in its extension anyway. If $e$ and $b$ are joined to generate $eb$, then we need to compare $b$ with the second item $f$ in $T^\prime$ ($e$ must match successfully when building $1$-itemsets). We get $sn(b)$ $<$ $sn(f)$. This means that item $f$ may match for extended itemsets later, thus this itemset needs to be preserved. In Fig. \ref{fig:kUL}, if $eb$ and $ef$ are joined to generate $ebf$, the match is successful.

\section{Performance Evaluation}
\label{sec:experiments}

In this section, we evaluate the performance of the THUIM algorithm  in detail and compares it with that of TargetUM \cite{miao2021targetum,miao2022targetum}. This latter constructs a TP-tree during the mining process, which makes it necessary to take this memory into account each time that the dataset is mined. Obviously, the minimum threshold ($\sigma$) is inversely proportional to the number of itemsets found, and there are fewer results when the threshold is larger. Therefore, when $\sigma$ is set to a relatively small value, TargetUM runs out of memory and the algorithm terminates. THUIM uses a pattern matching mechanism instead of searching for a complete result set, and it directly mines the target high-utility itemsets. To a certain extent, THUIM can discard itemsets that are not the target itemsets in advance, which improves its performance.

\subsection{Experimental setup}

Experiments were conducted on a PC having an AMD Ryzen 5 3600 6-Core Processor 3.60 GHz, 8 GB of memory and 64-bit Microsoft Windows 10. TargetUM and THUIM are implemented in Java. We make all source code available at GitHub {\color{blue} https://github.com/DSI-Lab1/TargetUM}

Six datasets were selected in this experiment, including four real-life datasets and two synthetic datasets, namely chess, retail, BMSPOS2, ecommerce, T10I4D100K, and T40I10D100K. For each dataset, some of its basic characteristics are summarized in a total of seven points, which are 1) the name of the dataset, 2) the size of the dataset, 3) the number of transactions contained in the dataset, 4) the number of distinct items  in the dataset, 5) the maximum length of transactions in the dataset, 6) the average length of transactions from the dataset, and 7) the total utility of the dataset. The information is presented in  Table \ref{Tab:dataset}. In this experiment, $\sigma$ is set to specific values, thus knowing the total utility of the dataset in advance can  help us determine the range of threshold values and  plays a very important role.

BMSPOS2, retail, chess and ecommerce are four real-life datasets.  BMSPOS2  contains several years of point-of-sale data from a major electronics retailer. Retail is transactions from a retail store. Chess is compiled from the UCI chess dataset, and ecommerce contains all transactions that occurred between January 12, 2010, and September 12, 2011 that took place in a UK registered online store. T10I4D100K and T40I10D100K are synthetic datasets generated by the IBM generator, which are relatively stable and are usually used to evaluate the scalability of algorithms. Note that for the external utility of the dataset, only ecommerce has real external utility values, and the external utility of other datasets are randomly generated.

\begin{table}[!htb]
	\centering	
	\caption {Characteristics of different datasets}
	\label{Tab:dataset}
	\begin{tabular}{l|c | c |c |c |c}
		\hline \hline
		
		\textbf{Dataset}  & \textbf{Trans} & \textbf{Items} & \textbf{MaxLen} & \textbf{AvgLen} & \textbf{TotalUtility} \\  \hline
		chess  & 3196   & 75  & 37  & 37.0 & 2,156,659 \\ 
		retail  & 88162 & 16468 & 76 & 10.3 & 362,481,272 \\ 
		BMSPOS2  & 515366 & 1656 & 164 & 6.5 & 1,301,704,112 \\ 
		ecommerce  & 14975 & 3468 & 29 & 11.6 & 49,701,3754 \\ \hline
		T10I4D100K  & 100000 & 870 & 29 & 10.1 & 388,548,246 \\ 
		T40I10D100K & 100000 & 942 & 77 & 39.6 & 1,550,463,496 \\
		\hline \hline
		
	\end{tabular}
\end{table}

\subsection{Efficiency analysis}

To evaluate the performance of THUIM,  it is compared with that of TargetUM. It is clear that TargetUM and THUIM produce the same results for the same dataset, minimum utility threshold ($\sigma$) value, and target pattern ($T^\prime$). Therefore, this point will not be the focus of this comparison but instead three aspects, which are the running time, memory consumption, and number of candidate itemsets that are generated. As far as we know, as a tree-based algorithm, TargetUM consumes a lot of memory when constructing the TP-tree. Considering the limitations of the experimental equipment, the total utility value of each dataset was calculated so as to set an appropriate utility threshold value that ensures the actionability of experimental results. At the same time for this experiment, the size of  $T^\prime$ is set to three to five items, for retail, $T^\prime$ = \{988, 990, 991, 998, 1003\}, for ecommerce, $T^\prime$ = \{21844, 23052, 23166, 8501412\}, for T10I4D100K, $T^\prime$ = \{85, 447, 859\}, for BMSPOS2, $T^\prime$ = \{11, 23, 36\}, for chess, $T^\prime$ = \{48, 66, 70, 72\}, and for T40I10D100K, $T^\prime$ = \{521, 872, 933\}.

\begin{figure*}[!htb]
	\centering
	\includegraphics[trim=120 0 90 0,clip,scale=0.425]{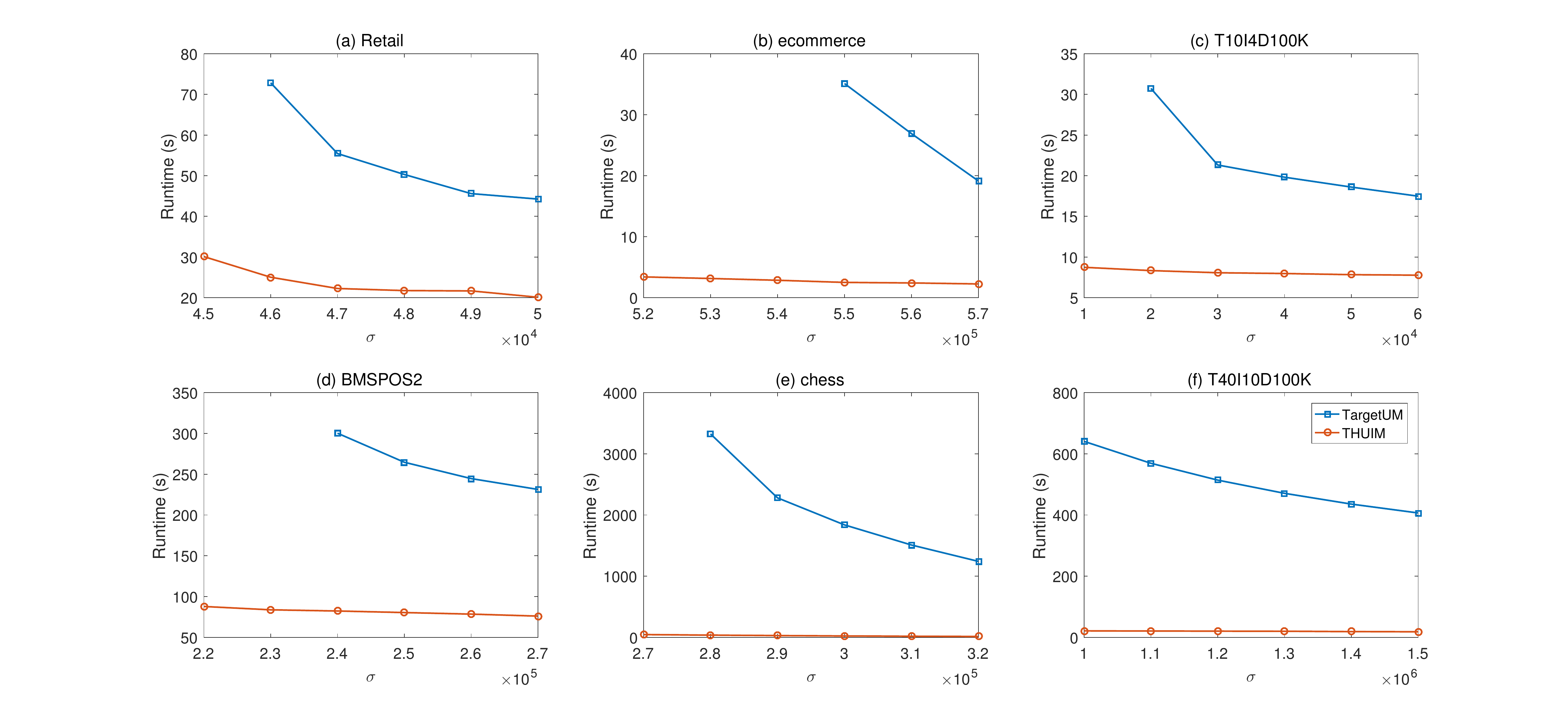}
	\caption{Runtimes for varied $\sigma$ values and a fixed $T^\prime$ (a) Retail ($T^\prime$ = \{988, 990, 991, 998, 1003\}). (b) ecommerce ($T^\prime$ = \{21844, 23052, 23166, 8501412\}). (c) T10I4d100K ($T^\prime$ = \{85, 447, 859\}). (d) BMSPOS2 ($T^\prime$ = \{11, 23, 36\}). (e) chess ($T^\prime$ = \{48, 66, 70, 72\}). (f) T40I10D100K ($T^\prime$ = \{521, 872, 933\})}
	\label{fig:runtime}
\end{figure*}

\begin{figure*}[!htb]
	\centering
	\includegraphics[trim=120 0 90 0,clip,scale=0.425]{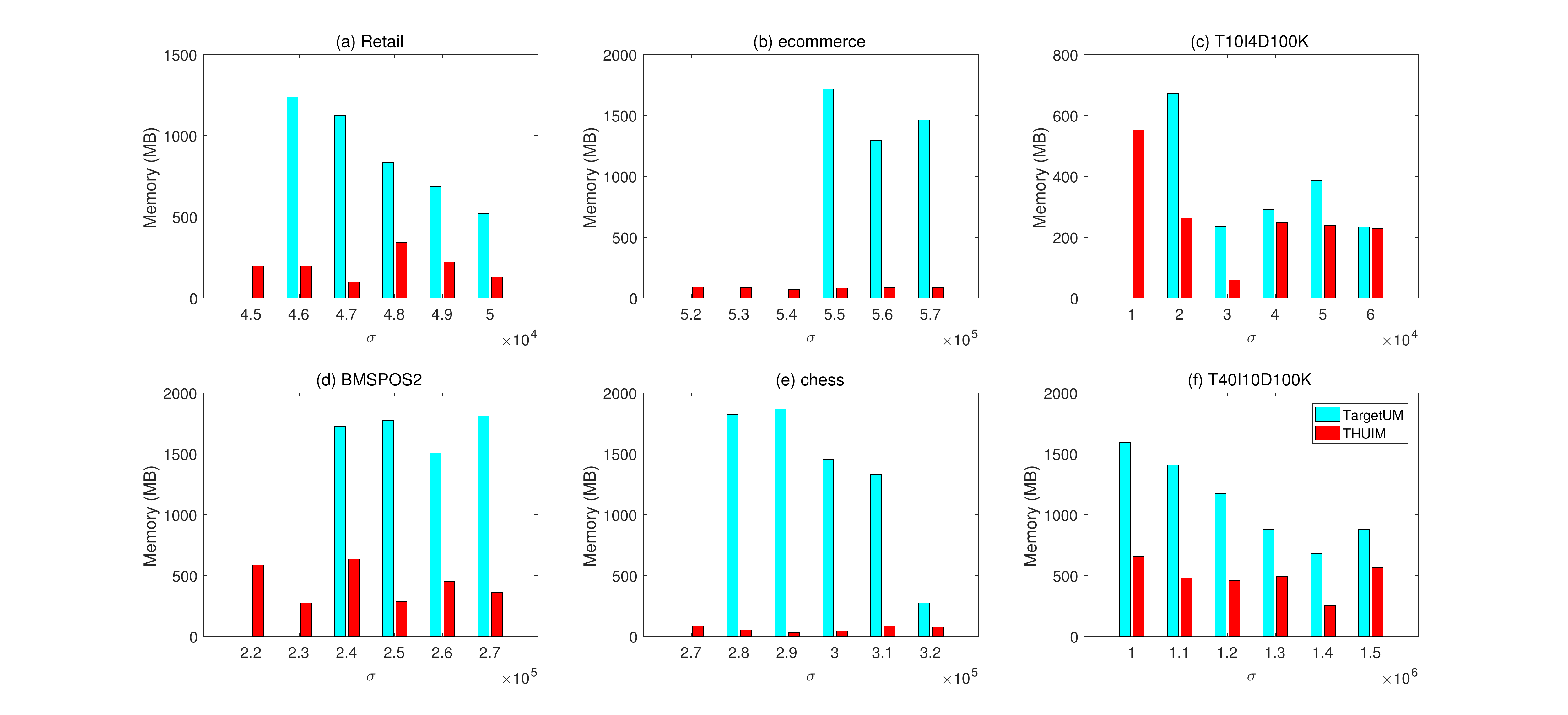}
	\caption{Memory when varied $\sigma$ is varied for a fixed $T^\prime$ (a) Retail ($T^\prime$ = \{988, 990, 991, 998, 1003\}). (b) ecommerce ($T^\prime$ = \{21844, 23052, 23166, 8501412\}). (c) T10I4d100K ($T^\prime$ = \{85, 447, 859\}). (d) BMSPOS2 ($T^\prime$ = \{11, 23, 36\}). (e) chess ($T^\prime$ = \{48, 66, 70, 72\}). (f) T40I10D100K ($T^\prime$ = \{521, 872, 933\})}
	\label{fig:memory}
\end{figure*}

Fig. \ref{fig:runtime} shows the running times of THUIM and TargetUM on six datasets. As can be seen from the figure, THUIM always has smaller running times than TargetUM, and runtimes decrease linearly as $\sigma$ is increased. There are two main reasons why THUIM has small runtimes. First, THUIM is more efficient than TargetUM as THUIM can mine the target high-utility itemsets directly, whereas TargetUM requires the construction of a TP-tree first. Second, to ensure that TargetUM can produce some results,  $\sigma$ is set to relatively large values. Despite that, it can be seen in Fig. \ref{fig:runtime} that for some datasets such as retail, ecommerce, T10I4D100K, BMSPOS2 and chess, TargetUM has no results for some threshold values. This is because TargetUM ran out of memory when constructing the TP-tree. The difference in running time between TargetUM and THUIM ranges from tens to thousands of times, which is due to the fact that TargetUM mines all high-utility itemsets that satisfy $\sigma$ when constructing the TP-tree, which wastes a lot of time. On the other hand, THUIM can directly discover the target high-utility itemsets $x$ using its pattern matching mechanism, which saves a lot of time.

Memory consumption was monitored in the experiment. Results are shown in Fig. \ref{fig:memory}. It can be seen that the memory consumption of THUIM and TargetUM do not increase linearly as the threshold value is decreased. However, from an overall perspective, the memory consumed by TargetUM is much larger than that of THUIM. This is due to the fact that when constructing utility-lists, THUIM with the pattern matching mechanism will filter out utility-lists that do not contain target items, thus reducing the construction of utility lists. On the other hand, TargetUM first finds the full set that satisfies $\sigma$, and then construct a large number of utility-lists. TargetUM consumes much memory to store the utility-lists to find the target high-utility itemsets. Therefore, THUIM has a great advantage in terms of memory consumption.

\begin{figure*}[!htb]
	\centering
	\includegraphics[trim=120 0 50 0,clip,scale=0.42]{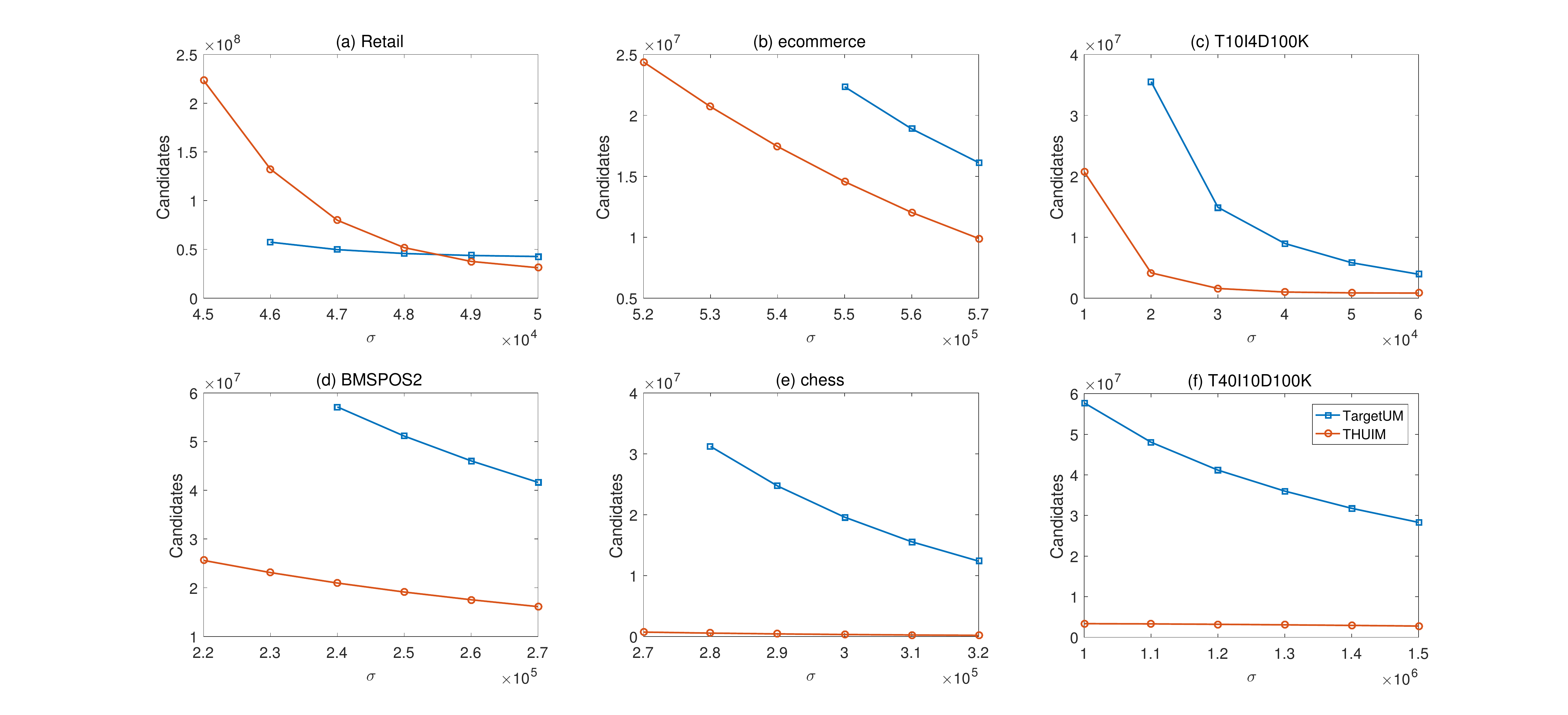}
	\caption{Candidates under varied $\sigma$ with a fixed $T^\prime$ (a) Retail ($T^\prime$ = \{988, 990, 991, 998, 1003\}). (b) ecommerce ($T^\prime$ = \{21844, 23052, 23166, 8501412\}). (c) T10I4d100K ($T^\prime$ = \{85, 447, 859\}). (d) BMSPOS2 ($T^\prime$ = \{11, 23, 36\}). (e) chess ($T^\prime$ = \{48, 66, 70, 72\}). (f) T40I10D100K ($T^\prime$ = \{521, 872, 933\})}
	\label{fig:candidate}
\end{figure*}

\begin{figure*}[!htb]
	\centering
	\includegraphics[height=0.17\textheight,width=1.05\linewidth,trim=100 0 50 0,clip,scale=0.42]{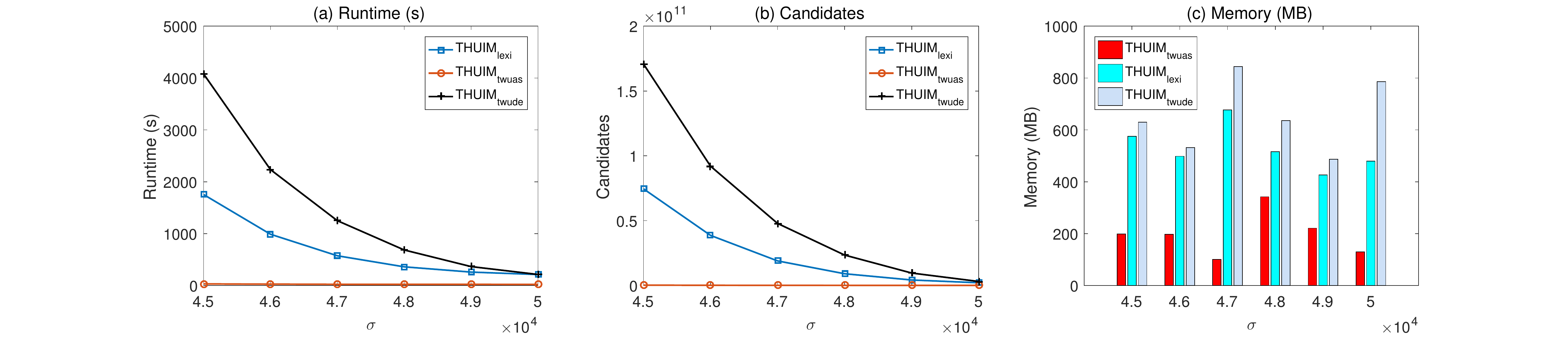}
	\caption{Different sorting strategies are compared for varied $\sigma$ values on the same retail dataset  and a fixed $T^\prime$ = \{988, 990, 991, 998, 1003\} (a) Runtime comparison. (b) Candidate comparison. (c) Memory consumption comparison}.
	\label{fig:scaorder}
\end{figure*}

When THUIM constructs a utility-list, the utility and the remaining utility are calculated at the same time. This process based on the HUI-Miner algorithm avoids multiple database scans to mine high utility itemsets in one phase. However, the construction of utility-lists can also be regarded as candidates. Thus, in a sense, the fewer utility-lists are constructed during the mining process, the more efficient the algorithm is. And the efficiency of the algorithm should be improved as much as possible while guaranteeing the completeness of the results. To further verify the efficiency of THUIM, the number of generated candidates was counted. It is shown in Fig. \ref{fig:candidate}. As can  be seen in that figure, the number of candidates gradually decreases, showing a linear change as $\sigma$ is increased. Besides, THUIM generates far less candidates than TargetUM, thanks to its pattern matching mechanism. Therefore, in summary, THUIM is more efficient  than TargetUM.

\subsection{Scalability analysis}

In previous experiments, we assessed the performance of THUIM on different datasets for different $\sigma$ values. Here, we further analyze the impact of the sorting strategy and  scalability for the number of transactions. For this experiment, we have selected retail, T10I4D100K and T40I10D100K as the test datasets, which are all sparse datasets and are appropriate for scalability experiments.

\textbf{Processing order of items}. The processing order of items for a dataset, which is used for the sorting strategy, plays a very important role in the mining process of THUIM. That is to say, choosing an appropriate sorting order can greatly improve the performance of the algorithm. THUIM adopts the \textit{TWU}-ascending order. It should be clear that in THUIM, the target pattern is also sorted according to that same order, so changing the sorting strategy has no effect on the output of THUIM. We therefore measured the effects of the sorting strategy on runtime, candidate generation and memory consumption, and compared three different sorting strategies, namely the \textit{TWU}-ascending order, the lexicographic order and the \textit{TWU}-descending order, denoted as THUIM$_{\textit{twuas}}$, THUIM$_{\textit{lexi}}$ and THUIM$_{\textit{twude}}$. Figure \ref{fig:scaorder} shows the experimental results on the retail dataset for a fixed $T^\prime$ = \{988, 990, 991, 998, 1003\}. As can be seen in that figure,  THUIM$_{\textit{twuas}}$ has  excellent performance. Besides, THUIM$_{lexi}$ performs better than THUIM$_{\textit{twude}}$. Experimental results show that choosing THUIM$_{\textit{twuas}}$ can greatly reduce the number of utility-lists built and  reduce runtime and  memory usage.

\begin{table*}[htb]
	\fontsize{8pt}{11pt}\selectfont
	\centering
	\caption{Characteristics of different subsets of T40I10D100K and T10I4D100K}
	\label{Tab:sizedatasize}
	\begin{tabular}{c|c|llllll}
		\hline \hline
		\multirow{2}{*}{\textbf{Dataset}} & \multirow{2}{*}{\textbf{Type}} & \multicolumn{6}{c}{\textbf{Characteristics}} \\ \cline{3-8}
		& & \textbf{50K} & \textbf{60K} & \textbf{70K} & \textbf{80K} & \textbf{90K} & \textbf{100K} \\ \hline
		
		\multirow{3}{*}{T40I10D100K} & Size (KB) & 15264 & 18309 & 21371 & 24425 & 27471 & 30521 \\ \cline{2-8}
		& Trans & 50000 & 60000 & 70000 & 80000 & 90000 & 100000 \\ \cline{2-8}
		& Total Utility & 775,628,384 & 930,254,499 & 1,085,647,233 & 1,240,566,686 & 1,395,442,914 & 1,550,463,496 \\ \hline
		
		\multirow{3}{*}{T10I4D100K} & Size (KB) & 4134 & 4964 & 5791 & 6617 & 7442 & 8269 \\ \cline{2-8}
		& Trans & 50000 & 60000 & 70000 & 80000 & 90000 & 100000 \\ \cline{2-8}
		& Total Utility & 194,119,368 & 233,062,036 & 271,856,612 & 310,634,977 & 349,562,923 & 388,548,246 \\ \hline \hline
	\end{tabular}
\end{table*}

\begin{figure*}[!htb]
	\centering
	\includegraphics[height=0.17\textheight,width=1.05\linewidth,trim=100 0 50 0,clip,scale=0.42]{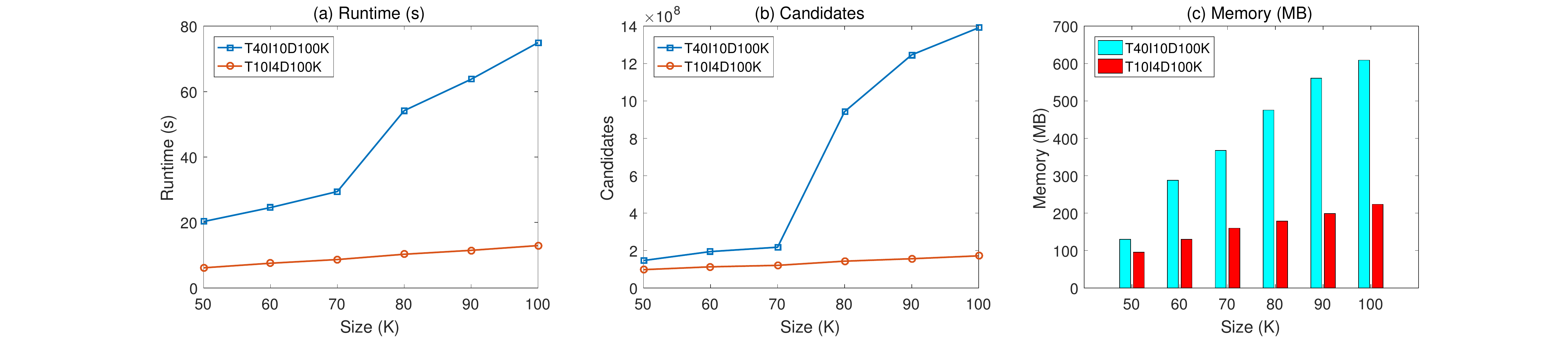}
	\caption{Influence of dataset size (with a 10K increment) for a fixed $\sigma$ (T10I4D100K $\sigma$ = 1000 and T40I10D100K $\sigma$ = 30000) and a fixed $T^\prime$ (T10I4D100K $T^\prime$ = \{85, 447, 859\} and T40I10D100K $T^\prime$ = \{390, 464, 515, 611, 922\}) (a) Runtime comparison. (b) Candidates comparison. (c) Memory consumption comparison}.
	\label{fig:scasize}
\end{figure*}

\textbf{Length of datasets}. To evaluate the scalability of THUIM, the datasets T10I4D100K and T40I10D100K were used, both of which have 100,000 transactions. For each dataset, six subsets were created, containing 10K (i.e., 10,000) more transactions than the previous one. The characteristics of these subsets with 50K, 60K, 70K, 80K, 90K, and 100K transactions are shown in Table \ref{Tab:sizedatasize}. For T10I4D100K,  parameters were set as  $\sigma$ = 1000 and $T^\prime$ = \{85, 447, 859\}, and for T40I10D100K, parameters are $\sigma$ = 30000 and $T^\prime$ = \{390, 464, 515, 611, 922\}. Experiments were carried out to evaluate three aspects: runtime, candidate generation, and memory consumption. The detailed experimental results are shown in Fig. \ref{fig:scasize}. As can be seen in Fig. \ref{fig:scasize}, the performance changes linearly as the amount of data (10K) increases. Besides, the setting of the target patterns has a certain impact on the performance of THUIM. In summary, compared with TargetUM, THUIM is a quite efficient algorithm, which can quickly and effectively discover the target high-utility itemsets.

\section{Conclusion}
\label{sec:conclusion}

This paper proposed the THUIM algorithm, which combines a target pattern with the HUI-Miner algorithm using a pattern matching mechanism. Targeted pattern mining is a recently proposed problem, and only the TargetUM algorithm was available. While TargetUM suffers from many limitations, and its performance is impaired by those especially for dense datasets and large datasets. This may result in problems such as out of memory errors and failing to produce the expected results. THUIM adopts a matching mechanism to obtain serial numbers for all items based on the \textit{TWU} order, so as to better perform itemset comparison during the mining process, and to filter the high-utility itemsets that do not satisfy the constraints of $T^\prime$ and $\sigma$ in advance, which can speed up the mining process. Experiments on different datasets and a comparison with the \textit{TargetUM} algorithm have shown that THUIM has advantages over TargetUM in terms of running time and memory, and it has good scalability. 

In the future, it is expected that the problem of targeted pattern mining will receive more interest from researchers and be applied in many fields such as for privacy protection, product recommendation, and intelligent search. Proposing a way to balance the relationship between an itemset and its match with a target pattern ($T^\prime$) and its utility w.r.t the utility threshold ($\sigma$)  is also worth considering. We also expect more efficient and excellent algorithms.

\ifCLASSOPTIONcaptionsoff
  \newpage
\fi

\bibliographystyle{IEEEtran}
\bibliography{THUIM}


\end{document}